# Photometric Deprojection of Edge-on Galaxies

## D. V. Bizyaev*

*Sternberg Astronomical Institute, Universitetskiĭ pr. 13, Moscow, 119899 Russia*
Received June 1, 1999; in final form, September 24, 1999

**Abstract**—Photometric deprojection is used to determine the stellar-disk and bulge parameters for several edge-on galaxies from the FGC catalog. The assumption that the galaxies of our sample belonging to the fourth (i.e., lowest) surface-brightness class in the FGC are edge-on, low-surface-brightness (LSB) galaxies is considered. © *2000 MAIK "Nauka/Interperiodica".*

## INTRODUCTION

The central brightness $\mu_0$ of the disks in most spiral galaxies varies over a relatively narrow range and is, on the average, $21–22^m$ in the photometric $B$ band [1]. However, some of the galaxies have disks with a considerably lower central surface brightness. Such objects were called low-surface-brightness (LSB) galaxies to distinguish them from high-surface-brightness (HSB) galaxies, whose $\mu_0$ do not differ markedly from $21–22^m$. McGaugh [2] suggested a more detailed classification of galaxies depending on $\mu_0$; however, the simple division into LSB and HSB objects is also widely used in the literature.

According to [3–5], the LSB disk galaxies are distinguished by their structure and evolution. The main properties of LSB galaxies include the following: low surface and volume densities of their stellar disks; large disk sizes compared to HSB galaxies; heavy-element underabundance; high gas mass-to-light ratios; a large extent of the rising portion of the rotation curve; and low star-formation rates. In contrast to HSB galaxies, dark halos can play a crucial role in the formation and evolution of LSB galaxies [3, 6].

The fact that there are low-surface-brightness objects among edge-on galaxies is mentioned in the description of the Flat Galaxy Catalogue (hereafter FGC [7]) and by Reshetnikov and Combes [8] (hereafter RC97). Karachentsev *et al.* [9] analyzed radial brightness profiles of 120 flat galaxies and showed some of them to have disks of low central surface brightness. After this paper had already been submitted and returned with the referee's comments, the preprint by Matthews *et al.* [10] with the results of a study of UGC 7321, an edge-on LSB galaxy, appeared in the SISSA database.

Reshetnikov and Combes [11] (hereafter RC96) and RC97 report the results of photometric studies for a number of edge-on galaxies. Several galaxies considered by RC97 were included in the FGC, and their inclinations do not differ significantly from 90°.

In this study, we consider a small sample of galaxies, some of which presumably have disks of low surface brightness. Since edge-on LSB galaxies have not yet been studied in detail, our goal here is to compare the disk and bulge parameters for the LSB and HSB galaxies from our sample. To determine these parameters, we use deprojection, which is based on modeling the surface-brightness distributions for the disks and bulges of edge-on galaxies.

## SAMPLE OF GALAXIES

We use the $B$, $V$ (Johnson) and $I$ (Cousins) images of galaxies obtained on July 21–23, 1993, with the 1.2-m telescope at the Observtatoire de Haute Provence (OHP) in France. The observations were performed by Reshetnikov and Combes [11]. The CCD detector, observing conditions, and image calibration were described in more detail by RC96. Reshetnikov kindly provided the galaxy images to the author. Table 1 gives: (1) the objects' names, (2) their distances for the Hubble constant $H = 75$ km s$^{-1}$ Mpc$^{-1}$, and (3) the FGC surface-brightness classes. Two and four galaxies have high and low surface brightnesses, respectively. One more galaxy, UGC 11132, was not included in the FGC; nevertheless, we performed photometric deprojection for it as well, because its inclination to the plane of the sky is also close to 90°.

We additionally removed cosmic-ray hits from the galaxy images and masked field stars. The galaxy position angles were determined from the condition of maximum axial symmetry.

We measured the total colors of the galaxies within an ellipse with the semimajor axis equal to 0.9 of the optical radius $R_{25}$ taken from the RC3 catalog [12]. The semiminor axis was assumed to be 0.1 of the semimajor axis, which is close to 1/7, the limiting axial ratio for a galaxy to be included in the FGC.

---

* E-mail address for contacts: dmbiz@sai.msu.su

**Table 1**

| UGC | $D$, Mpc | Surface-brightness class |
|-----|----------|--------------------------|
| 11132 | 37.9 | – |
| 11230 | 94.7 | 4 |
| 11301 | 60.0 | 3 |
| 11838 | 46.3 | 2 |
| 11841 | 79.9 | 4 |
| 11859 | 40.2 | 4 |
| 11994 | 65.0 | 2 |

We computed the $B$–$V$ and $V$–$I$ colors of the galaxies by using calibrations from RC96. The total $B$–$V$ colors uncorrected for extinction in the galaxies themselves differ, on the average, by $0.^{m}10$ from those in RC97. At the same time, the mean $V$–$I$ color difference is less significant, $0.^{m}007$. These discrepancies can be explained by different procedures of allowance for the contribution of field stars to the total galaxy color and by the fact that the galaxy colors were determined by RC97 using individual calibration relations for each night, which slightly differ from the average ones in RC96.

UGC 11841 is an exception. For this object, the discrepancies between our results and those of Reshetnikov and Combes are large, $0.^{m}40$ and $0.^{m}38$ for $B$–$V$ and $V$–$I$, respectively. The source of this uncertainty is a bright field star located a mere 8″ from the galaxy center and superimposed on its image.

To determine the $B$–$V$ and $V$–$I$ color indices separately for the galaxy stellar disks, we masked and excluded regions located within two pixels of the galactic plane and regions adjacent to the galaxy center within 1.5 of the optical disk scale. Figure 1 shows the total color indices of the galaxies in comparison with the colors of their disks. The colors were corrected for Galactic extinction using values from the LEDA database (Lyon Observatory). As is evident from the figure, the $B$–$V$ and $V$–$I$ color indices of the stellar disks are, on the average, $0.^{m}2$ smaller; i.e., they are bluer than the galaxies as a whole. The filled triangles indicate galaxies of the fourth surface-brightness class, and the open triangles indicate objects of the second and third classes, as well as UGC 11132. No systematic differences between the colors of galaxies of different surface brightness classes are observed. Our sample contains galaxies with disk colors typical both of normal HSB galaxies (see, e.g., [13]) and LSB galaxies with large radial disk scales [14, 15]. The LSB galaxies are generally believed to be bluer. However, according to [16], only the color indices of small LSB galaxies differ significantly from those of HSB galaxies.

## PHOTOMETRIC DEPROJECTION OF EDGE-ON GALAXIES

The photometric parameters of the disks and bulges were determined by fitting model components (disk and bulge) to the observed two-dimensional $I$ brightness distribution.

The galaxy region used for the fitting was bounded by an ellipse whose major axis was oriented along the previously inferred position angle. The ellipse semiaxes were chosen according to the above criterion ($a \approx 0.9R_{25}$ and $b \approx 0.1a$).

For the model disk, we specified the distribution of volume luminosity density by an exponential law of the form

$$\rho_d(r, z) = \rho_{0d} e^{-\frac{r}{R_e}} e^{-\frac{z}{Z_e}}, \qquad (1)$$

where the radial and vertical scales $R_e$ and $Z_e$ and the central luminosity density $\rho_{0d}$ are the model parameters of the problem.

Since all the objects studied were included in the Flat Galaxy Catalogue, we assume that they are seen exactly edge-on. The dust lane in the images is not displaced significantly relative to the galaxy plane. The symmetry of the image about the galaxy plane also provides evidence that the galaxy inclination deviates from 90° by no more than a few degrees.

We obtained the model disk image by integrating equation (1) along the line of sight. The radial extent of the disk was assumed to be 1.3 $R_{25}$, which roughly corresponds to four disk scales $R_e$ (see, e.g., [17]).

To eliminate the effect of dust absorption on the actual surface brightness distribution, we did not consider regions located in the vicinity of the galaxy plane and masked regions within ±2 pixels of it. For the assumed distances to the objects (see Table 1), this is equivalent to excluding regions within 440 pc of the galaxy plane, on the average. We assumed dust absorption to be insignificant in regions outside the two excluded pixels.

When comparing the model disk brightness profile with the observed one, we also excluded the central part of the disk ($R < R_e$), where the bulge can contribute significantly to the luminosity.

The fitting was performed by minimizing the sum of the squares of deviations of the model profile from the observed one. We found the photometric disk parameters thus obtained to depend only slightly on whether the weighting function was used for the minimization or not.

The conversion to surface brightness (in magnitudes) was made by using calibration relations from RC96.

To determine the bulge parameters, we subtracted the model disk from the observed image. The central part of the remainder was then fitted by the brightness distribution

$$I_b(X, Y) = \frac{I_{b0}}{1 + \dfrac{X^2 + Y^2}{R_{King}^2}}, \qquad (2)$$

where $X$ and $Y$ are the coordinates in the plane of the sky. The central surface brightness $I_{b0}$ and the characteristic bulge radius $R_{King}$ are the model parameters. Apart from bulges whose surface brightness was specified by King's law (2), we considered de Vaucouleurs's law of brightness variations. In this case, however, the frequent divergence during minimization forced us to prefer the law (2). Large-scale images are required to infer the galaxy bulge parameters with a high accuracy. The available images allow the integrated bulge parameters, i.e., their total magnitudes and luminosities, to be estimated without substantial errors.

## RESULTS OF DEPROJECTION

Table 2 contains the parameters inferred by deprojection. The table gives the galaxy name, the radial disk scale for the distance in Table 1, the vertical disk scale, the scale ratio $Z_e/R_e$, and the $I$ central disk surface brightness $\mu_{d0}$ reduced to an inclination of 0°. Table 2 also gives the $I$ central bulge surface brightness $\mu_{b0}$ (in magnitudes) and the King radius $R_{King}$ of the bulge.

The central disk and bulge surface brightness for face-on galaxies was determined by using calibrations from RC96 and corrected for Galactic extinction by using data from the LEDA database.

The seeing was ~2–3″ during the observations. According to RC96, allowance for atmospheric effects when comparing images with the model causes the disk parameters to change by no more than 20%.

To estimate the errors in the quantities listed in Table 2, we formed several model images composed of disks and bulges whose luminosities were given by formulas (1) and (2). We added the noise typical of the images of our sample galaxies to the model brightness distributions. The parameters of the images formed in this way were obtained by using the deprojection procedure described above. The typical range of admissible photometric disk and bulge parameters was found to be no larger than 10%.

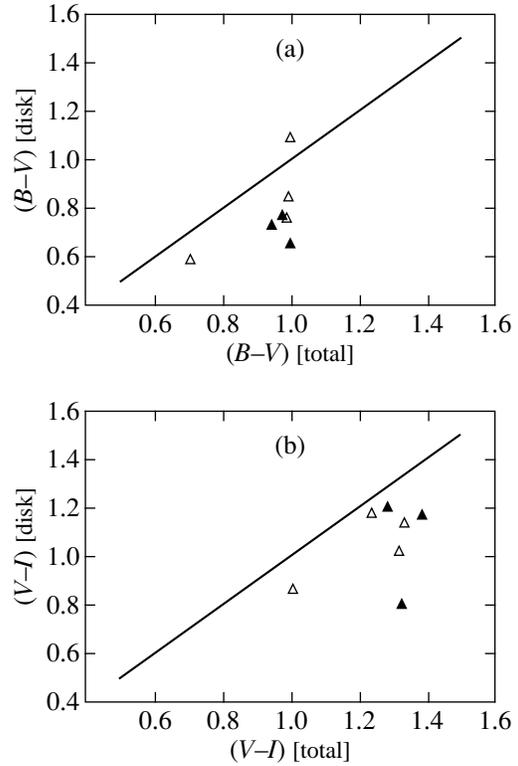

Fig. 1. Total color indices of the galaxies, $B$–$V$ (a) and $V$–$I$ (b), in comparison with the colors of their disks. The colors were corrected for Galactic extinction. The filled triangles represent galaxies of reduced surface brightness belonging to the fourth surface-brightness class in the FGC, and the open triangles indicate all the remaining galaxies of the sample. The straight lines correspond to equality of the color indices for the disks and galaxies as a whole.

To compare our radial and vertical disk scales with those inferred by RC97, who used individual cuts (without modeling), we considered not only the exponential distribution of volume luminosity (1), but also the

**Table 2**

| UGC | $R_e$, kpc | $Z_e$, kpc | $Z_e/R_e$ | $\mu_{d0}$ | $\mu_{b0}$ | $R_{King}$, kpc |
|------|------|------|------|------|------|------|
| 11132 | 2.8 | 0.67 | 0.24 | $20^m.9$ | $21^m.7$ | 0.3 |
| 11230 | 8.2 | 1.96 | 0.24 | 21.6 | 23.3 | 0.6 |
| 11301 | 8.7 | 1.37 | 0.16 | 20.9 | 23.0 | 0.2 |
| 11838 | 4.1 | 0.82 | 0.20 | 21.3 | 26.9 | 0.1 |
| 11841 | 11.7 | 1.58 | 0.14 | 21.8 | 22.3 | 0.8 |
| 11859 | 3.2 | 0.51 | 0.16 | 22.2 | 20.9 | 0.7 |
| 11994 | 4.3 | 0.95 | 0.22 | 19.7 | 23.6 | 0.2 |

**Table 3**

| UGC | This paper | | | RC96 | | |
|------|------|------|------|------|------|------|
| | $R_e$, kpc | $Z_e$, kpc | $Z_e/R_e$ | $R_e$, kpc | $Z_e$, kpc | $Z_e/R_e$ |
| 11132 | 2.72 | 0.84 | 0.31 | 2.52 | 0.60 | 0.24 |
| 11230 | 7.00 | 2.44 | 0.35 | 8.96 | 1.47 | 0.16 |
| 11301 | 8.47 | 1.49 | 0.18 | 8.93 | 0.47 | 0.05 |
| 11838 | 3.86 | 1.05 | 0.27 | 5.52 | 0.85 | 0.15 |
| 11841 | 12.52 | 1.89 | 0.15 | 13.39 | 1.86 | 0.14 |
| 11859 | 3.46 | 0.65 | 0.19 | 2.97 | 0.45 | 0.15 |
| 11994 | 4.26 | 1.39 | 0.33 | 4.06 | 1.16 | 0.29 |

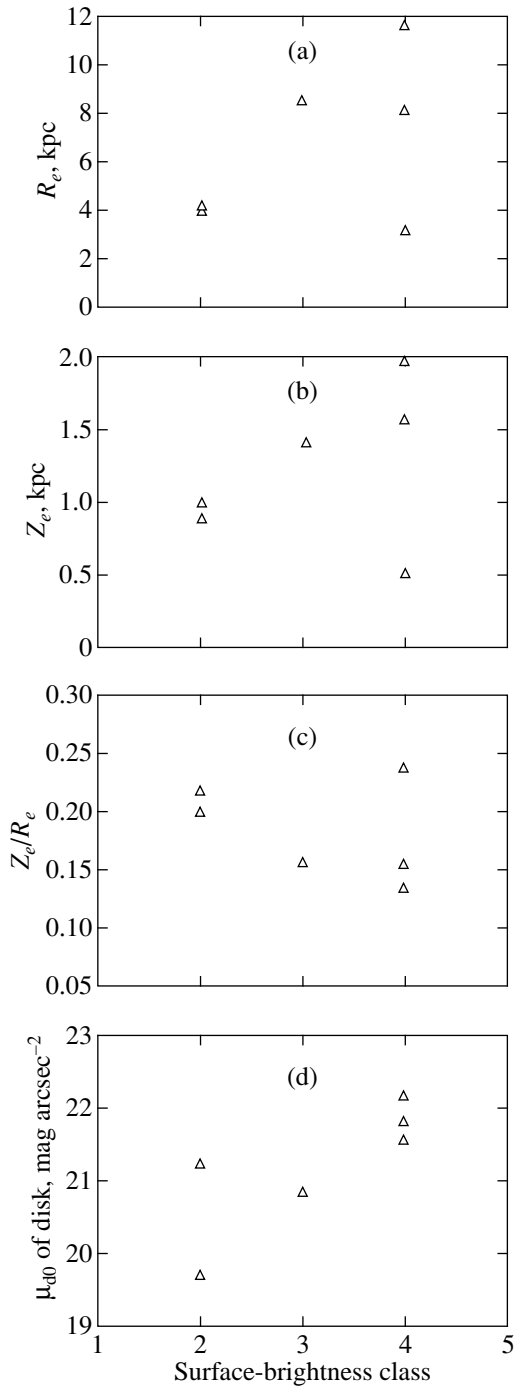

**Fig. 2.** The parameters of the model galaxy disks determined by fitting the *I*-band images. Radial scale $R_e$ (a), vertical scale $Z_e$ (b), $Z_e/R_e$ ratio (c), and central disk surface brightness $\mu_{d0}$ (d) are plotted against FGC surface-brightness class.

sech$^2\left(\dfrac{z}{Z_e}\right)$ vertical distribution. Table 3 gives $R_e$, $Z_e$, and $Z_e/R_e$ for this case in comparison with the results from RC97. The radial scales match closely, whereas our vertical scale $Z_e$ and $Z_e/R_e$ ratio are systematically higher. This may be because RC97 corrected their $Z_e$ for atmospheric effects; however, allowance for these effects causes $Z_e$ to decrease by no more than 20% (according to RC97). Note also that we determined the photometric parameters for our sample objects in the same way, which allows us to compare the surface brightnesses $\mu_{d0}$ and the $Z_e/R_e$ scale ratio of the disks.

To compare the results obtained by the method used here with the data obtained from individual cuts, we determined the disk parameters for the edge-on galaxy NGC 4244. The *I*-band image of this object was taken on February 13, 1994, with the JKT telescope at Observatorio del Roque de los Muchachos (La Palma). The exposure time was 1000 s. The image was provided via the Internet data archive by the Isaac Newton Group. The galaxy image was cleaned from cosmic-ray hits and flat-fielded. The photometric disk parameters were determined by using the procedure described above. The derived radial and vertical disk scales ($R_e = 104''$, $Z_e = 20''.3$) agree well with those obtained in [18] ($R_e = 107''$ and $R_e = 23''.9$, respectively) by analyzing individual cuts of the galaxy image taken on photographic plates.

Figure 2 shows plots of the galaxy disk parameters inferred by deprojection—$R_e$ (Fig. 2a), $Z_e$ (Fig. 2b), $Z_e/R_e$ (Fig. 2c), and the central surface brightness of the galaxy disks $\mu_{d0}$ (Fig. 2d)—against surface-brightness class. It is evident from Fig. 2d that the disk surface brightness for the fourth-class galaxies is about $1^m.5$ lower than that for the second-class galaxies. This is in good agreement with the division of galaxies into HSB and LSB, which differ by about $1^m.5$–$2^m$ in surface brightness (see, e.g., [19]).

Figures 2a and 2b suggest that there is no systematic difference between the radial and vertical disk scales of our sample galaxies belonging to different surface-brightness classes. The vertical-to-radial scale ratio tends to slightly decrease as we pass to galaxies of reduced surface brightness.

Thus, three galaxies of the fourth surface-brightness class in our sample have disks with reduced central surface brightness, which allows us to classify them as LSB galaxies. The disk scale of one of these galaxies (UGC 11841) exceeds 10 kpc, typical of large Malin 1-type galaxies (the so-called giant LSB galaxies; see [20]). Remarkably, UGC 11814 shows the lowest value of $Z_e/R_e \approx 0.136$ for our sample, in good agreement with that expected for flat edge-on galaxies [17].

One of the objects of our sample, UGC 11301, belongs to the third surface-brightness class and has a red disk (the $B$–$V$ color index of the disk is about $1^m.10$). The value of $\mu_{d0}$ for it does not exceed appreciably that found for the second-class galaxies. However, since several faint field stars are superimposed on the UGC 11301 image, the derived disk surface brightness is overestimated and represents a lower limit. The radial disk scale for UGC 11301 is 8.7 kpc, which is more typical of large galaxies with reduced surface brightness. This was also pointed out by RC97. In this case, we may be observing an object resembling the so-called red LSB galaxies. According to [21], such objects must also be encountered among LSB galaxies and account for up to 20% of their total number.

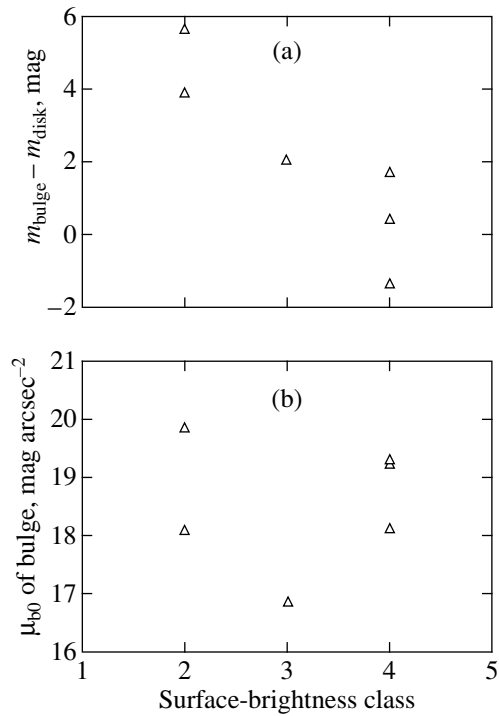

**Fig. 3.** Total bulge-to-disk luminosity ratio (in magnitudes) (a) and central bulge surface brightness (b) versus surface-brightness class.

Optical rotation curves have been published for three galaxies of our sample: UGC 11132, 11230, and 11944 [22, 23]. They all have an extended rising portion. However, strong extinction in edge-on galaxies distorts the shapes of their rotation curves, and it would be improper to directly compare their shapes on the basis of optical observations.

Figures 3a and 3b show the magnitude differences between the bulge and the disk, as well as the central bulge surface brightness, for different surface-brightness classes. In this case, the bulge parameters were determined by assuming King's brightness distribution (2). As is evident from Fig. 3, the LSB galaxies of our sample have bulges of higher relative luminosities. At the same time, the central bulge surface brightness does not depend on the mean disk surface brightness.

For UGC 11132, 11859, and 11994, the remainders obtained by subtracting the model disk and bulge from the observed brightness distribution suggest the presence of additional components in the central part of the galaxy. These galaxies probably have a bar or a lens that are not taken into account in the model. However, for these components to be studied in more detail, high-resolution galaxy images must be analyzed.

## CONCLUSION

The galaxies of our sample belonging to the fourth surface-brightness class according to the FGC have a number of properties typical of large LSB galaxies, namely, reduced central disk surface brightness and large radial disk scales.

The galaxies studied also show tendencies for the relative and absolute bulge luminosity to increase and for the vertical-to-radial scale ratio to decrease with decreasing mean surface brightness of the galaxy.

Our sample suggests that the FGC galaxies of the fourth surface-brightness class are mostly edge-on LSB galaxies. In that case, at least 3% (149 of 4455) of the FGC galaxies are candidates for galaxies of reduced surface brightness.

The FGC catalog may thus contain a considerable number of LSB galaxies with inclinations close to 90°. Their study will make it possible to reach conclusions about differences in the structure and evolution of the disk and spherical components of HSB and LSB galaxies.

## ACKNOWLEDGMENTS

I am grateful to Dr. V.P. Reshetnikov, who provided the galaxy images obtained at the Observatoire de Haute Provence, and to Prof. A.V. Zasov for a discussion and critical remarks, which helped supplement and improve the content of the paper. I am also grateful to the referee for valuable remarks, which helped improve the presentation of material and supplement the content of the paper. The photometry for the galaxy NGC 4244 was found using the Internet data archive operated by the Isaac Newton Group (La Palma). This study was supported by the Russian Foundation for Basic Research (project no. 98-02-17102).

## REFERENCES

1. P. van der Kruit, Astron. Astrophys. **173**, 59 (1987).
2. S. McGaugh, Mon. Not. R. Astron. Soc. **280**, 337 (1996).
3. E. De Blok, S. McGaugh, and T. van der Hulst, Bull. Amer. Astro. Soc. **189**, 8402 (1996).
4. G. Bothun, C. Impey, and S. McGaugh, Publ. Astron. Soc. Pacif. **109**, 745 (1997).
5. T. Pickering, C. Impey, J. van Gorkom, and G. Bothun, Bull. Amer. Astron. Soc. **184**, 1208 (1994).
6. W. de Blok and S. McGaugh, Astrophys. J. Lett. **469**, L89 (1996).
7. I. Karachentsev, V. Karachentseva, and S. Parnovsky, Astron. Nachr. **314**, 97 (1993).
8. V. Reshetnikov and F. Combes, Astron. Astrophys. **324**, 80 (1997).
9. I. Karachentsev, Ts. Georgiev, S. Kaĭsin, *et al.*, Astron. Astrophys. Trans. **2**, 265 (1992).

10. L. Matthews, L. Gallagher, and W. van Driel, astro-ph/9909142 (to be published in Astron. J.).

11. V. Reshetnikov and F. Combes, Astron. Astrophys., Suppl. Ser. **116**, 417 (1996).

12. G. de Vaucouleurs, A. de Vaucouleurs, H. Corwin, *et al.*, *Third Reference Catalogue of Bright Galaxies* (Springer, New York, 1991).

13. R. De Jong and P. van der Kruit, Astron. Astrophys., Suppl. Ser. **106**, 451 (1994).

14. J. Vennik, G. Richter, W. Thannert, *et al.*, Astron. Nachr. **317**, 289 (1996).

15. D. Sprayberry, C. Impey, G. Bothun, *et al.*, Astron. J. **109**, 558 (1995).

16. S. McGaugh and G. Bothun, Astron. J. **107**, 530 (1994).

17. A. Zasov, D. Makarov, and E. Mikhaĭlova, Pis'ma Astron. Zh. **17**, 884 (1991).

18. P. van der Kruit and L. Searle, Astron. Astrophys. **95**, 105 (1981).

19. R. Tully and A. Verheijen, Astrophys. J. **484**, 145 (1997).

20. T. Pickering and C. Impey, Bull. Amer. Astron. Soc. **186**, 3907 (1996).

21. J. Gerritsen and W. De Blok, Astron. Astrophys. **342**, 655 (1999).

22. I. Karachentsev, Pis'ma Astron. Zh. **17**, 485 (1991).

23. D. Makarov, I. Karachentsev, N. Tyurina, *et al.*, Pis'ma Astron. Zh. **23**, 509 (1997).

*Translated by A. Dambis*